# Moisture Effects on Nanowear of Gold Films

Xiaolu Pang[1,2], Alex A. Volinsky[2*], Kewei Gao[1]

[1]Department of Materials Physics and Chemistry, University of Science and Technology Beijing, Beijing 100083, China

[2]Department of Mechanical Engineering, University of South Florida, Tampa FL 33620, USA

**Abstract:** Nanowear properties of sputtered Au films in dry and wet environments were investigated using a scanning nanoindenter. Gold exhibits over 10 times higher wear rate in water compared to air at the same normal load of 10 µN. The friction coefficient obtained from scratch tests remained constant at 0.2 regardless of the testing conditions. Au surface roughness increased from 3 to 8 nm after 200 wear cycles in air. Surface ripples, 200 nm high developed on the Au film surface after 200 wear cycles in water. Film scratch hardness compares well with the nanoindentation hardness.

**Introduction**

Since gold is chemically inert and has relatively high melting point, it is used in electronic connectors [1], dental materials [2]-[6], jewelry [7], coinage [8], and so on. In many applications gold is in the form of a thin film or coating to minimize the cost. In these applications the mechanical properties such as friction, wear and adhesion are important and need to be investigated in different service environments. Saunders *et. al.* compared different ceramics and gold wear against human enamel in distilled water. They found that gold is the least abrasive material and is most resistant to wear [2]. The fact that gold produced less enamel wear than dental ceramics has also been reported [3]-[5]. Miyakawa *et. al.* discovered that gold and gold alloys exhibit good wear and lubrication properties even at high temperatures in vacuum [9]. In mechanical, dental and jewelry applications, gold wear happens at different scales in different environments because of the contact with fluids. This requires detailed characterization at the nano-micro-macro levels in order to relate the gold film tribology to adhesion, interface formation, microstructure changes, film thickness, and the substrate topography [1]. In many devices gold thin films are incorporated to create functionality, which highly depends on the materials properties, which in turn are a function of composition and microstructure. Gold does not chemically react with water, so chemical effects can be ruled out in the wear tests of gold, opposite to most other materials.

---

* Corresponding author: Alex Volinsky, Email address: volinsky@eng.usf.edu

The ability to resolve thin film deformation and failure modes is critical in determining device reliability and lifecycle. Wear resistance is one of the important mechanical properties in MEMS applications. Therefore, many techniques have been developed for evaluating micro and nano-friction and wear resistance, including those based on the atomic force microscopy [10]. The atomic force microscope (AFM) invention enabled measuring the material surface mechanical response at the atomic scale. Friction has been studied with AFM, and in recent years a strong influence of humidity on friction measurements was demonstrated [11]-[15]. There are fundamental differences between conventional large-scale friction measurements and those performed at the micro and nano-scales.

The scratch test is a simple and widely used method for investigating thin films adhesion, but is seldom used for hardness and wear resistance testing. Nanoindentation is another powerful technique for measuring mechanical properties of both bulk solids and thin films. Gold films are used in various environments; therefore, investigating their mechanical response in different working conditions is necessary for understanding film reliability. This paper reports the effects of moisture on friction, wear resistance, scratch and nanoindentation hardness of gold films at the nanoscale.

**Experiments**

Thin Au films were RF sputter-deposited onto 4" oxidized (100) silicon substrates to a nominal thickness of 3 μm, using Perkin Elmer 2400 RF sputtering system. A 15 seconds pre-sputter was used to prepare the substrate surface. The system was pumped down to 1 μTorr, after which Au films were deposited at 1000 W power. The maximum temperature during the sputter deposition reached 90 °C.

Hysitron Triboindenter equipped with either Berkovich or conical diamond indenter (1 μm tip radius) was used for wear, scratch and indentation tests. During wear testing, a normal force of 10 μN was applied to the diamond tip while it was moving over the sample surface. Wear tests were performed in ambient atmosphere (50% relative humidity) and in distilled water.

**Results and Discussion**

Fig. 1 shows the wear pattern of a gold film after 200 wear cycles under 10 μN normal load in air. Surface roughness of the worn area exhibits a significant change after 200 wear cycles, increasing from 3 to 8 nm. The wear mechanism of Au films in air is by grains pull-off, so some holes in the wear area can be seen; therefore, surface roughness increased with the number of wearing cycles. The average removal rate is about 0.07 nm of the film thickness per cycle in air vs. 1 nm in water at the same normal load of 10 μN, as seen in Fig. 2. Larger volume loss is noticeable in water compared to air; it is about 9 μm$^3$ after 100 wear cycles in water vs. only 1.4 μm$^3$ after 200 wear cycles in air. Possible reason of the significant wear rate increase in water is moisture-assisted fracture and plastic deformation of the abrasive grains, which otherwise would have carried major part of the applied load.

The ripple pattern topography that resulted from 200 surface scans in water with a sharp Berkovich tip is presented in Fig. 3. The modified area of 10x10 $\mu m^2$ is covered with periodic surface ripples aligned perpendicular to the scan direction. The width and depth of these ripples is quite uniform. Square area modified by the tip is surrounded by a frame of accumulated piled-up gold. This pattern on gold could not be reproduced in air, and was only observed during the water wear tests. Tip-induced wear and surface modification has been previously explored in various materials at the atomic and nano-scales [16]-[22]. We observed similar surface rippling of the KBr single crystal surface in air [23]. Once ripples are completely developed, their depth and periodicity increases with the number of consequent scans. This is a first report of the ripples forming in nanocrystalline materials vs. single crystals.

In 1950 Bowden and Tabor assumed that the lateral friction force, $F_L$, is proportional to the real contact area, $A$, and the shear strength, $\tau$ (mean lateral force per unit area):

$$F_L = \tau A \qquad (1).$$

If the shear strength is independent of pressure, the friction force is simply proportional to the tip contact area [24]. Fig. 4 shows the friction coefficient of gold films at various normal loads in different environments. The friction coefficient of 0.2 does not change much with increased normal load in air and in distilled water. Many papers report that humidity has a remarkable effect on the friction coefficient of various materials at the nano-scale [25]-[28]. It was found that the friction coefficient of mica shows a maximum at around 50% humidity, whereas the friction coefficients of $MoS_2$ and $Al_2O_3$ exhibit only slight variation with humidity [27]. Binggeli *et. al.* reported that the friction coefficient decreased with increasing humidity for silicon dioxide, but not for amorphous carbon [25]. The shear force and hydrophilic surface nature play an important role in the friction coefficient change. Studying gold tribological properties in water is fundamentally different compared to other materials, as chemical effects can be ruled out.

Wear resistance normally scales with the coating plastic properties, namely hardness or strength. One can obtain the tip contact area from the scratch track morphology and calculate the shear strength knowing the normal applied force during a scratch test. Gold films shear strength was calculated using equation (1). Fig. 5 shows the scratch track topography and the lateral friction force at 500 $\mu N$ normal applied load in air. The scratch track width is about 1.3 $\mu m$, which corresponds to 0.43 $\mu m^2$ contact area based on the tip geometry, shown schematically in Fig. 5c. With the friction lateral force of 90 $\mu N$ one would calculate the shear strength of about 209 MPa. The shear strength is simply half of the film yield stress. Based on our previous work, the yield stress of this gold film at room temperature is about 450 MPa [29], so the shear strength is 225 MPa, which is close to the 209 MPa value calculated from the scratch test results.

Scratch hardness can also be calculated from the scratch track topography and the normal applied load. The advantage of assessing the scratch hardness relative to indentation hardness is the ability to consider hardness variation along the scratch length and depth. The absolute hardness values, *H*, at specific positions can be calculated by measuring the groove width, *b*:

$$H = \frac{8F_N}{\pi b^2}$$ (2),

where $F_N$ is the normal applied load and *b* is the scratch track width shown in Fig 5c [31]. The value of gold film scratch hardness ranges from 632 MPa to 738 MPa in air. Fig. 6 shows the indentation hardness of gold film in air between 750 MPa to 1.1 GPa at different contact depth. Scratch and indentation hardness are in good agreement for this gold film in air at room temperature.

Plastic and wear properties are related, as normally harder materials have higher wear resistance. Gold is much softer than ceramic materials, although it has better wear resistance than hard ceramic materials [2-5]. This difference is due to the wear mechanisms operating at different scales and wear testing conditions.

**Conclusions**

Moisture significantly increases gold film wear rate due to a number of factors. Most important is the moisture-assisted fracture and plastic deformation of the abrasives which results in higher material removal rate [30].

Nano-ripples were formed on the gold surface after two hundred wear cycles in water. This phenomenon demonstrates that ripples can be achieved not only in single crystals, but in nanocrystalline materials as well.

The shear strength of the gold film is about 209 MPa, in good correlation with our former results of 225 MPa obtained from nanoindentation [29]. The gold film scratch hardness is between 632 and 738 MPa in air.

**Acknowledgements**

This work was supported by the National Natural Science Foundation of China (No. 50471091). Xiaolu Pang would like to acknowledge the support from the State Scholarship Fund of China (No. 20063037), and Alex Volinsky would like to acknowledge the support from NSF (CMMI-0600266, CMMI-0600231, and CMMI-0631526). Gold films were sputtered at Sandia National Laboratories in Livermore, CA.

**Figures**

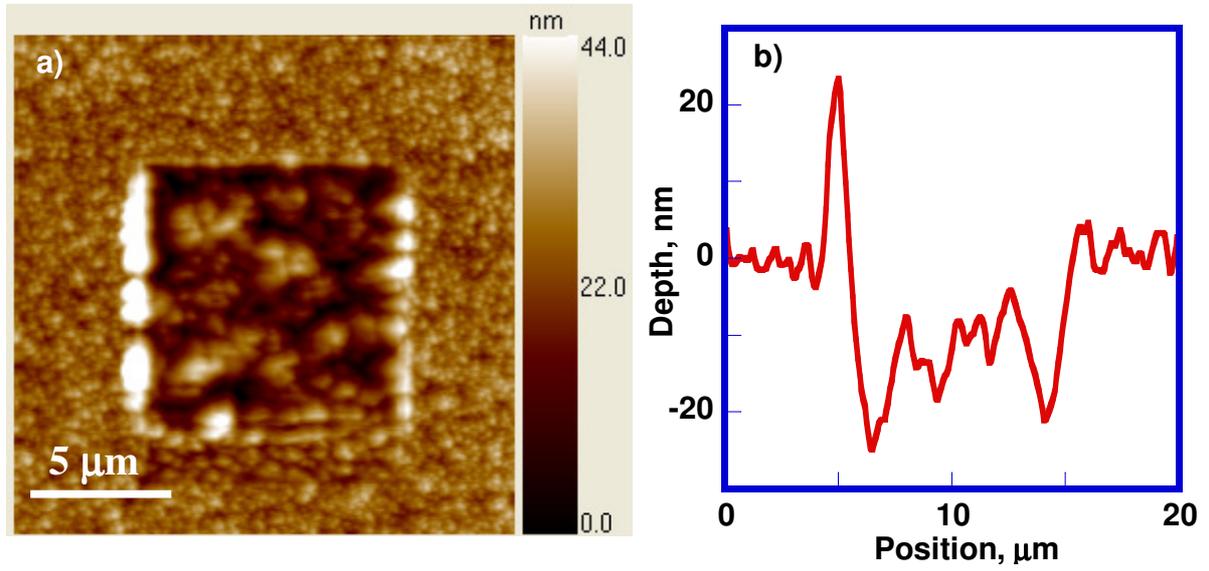

*Fig. 1. a) Wear area and b) wear track topography of a gold film after 200 wear cycles at 10 μN normal load with a blunt conical tip in air.*

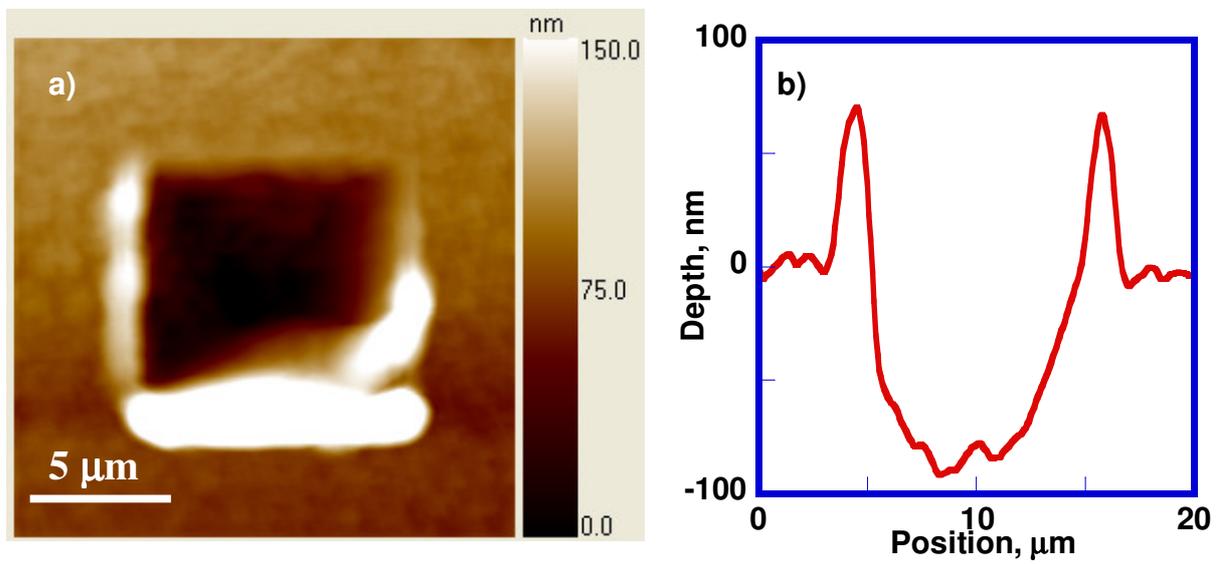

*Fig. 2. a) Wear area and b) wear track of a gold film after 100 wear cycles at 10 µN normal load with a blunt conical tip in water.*

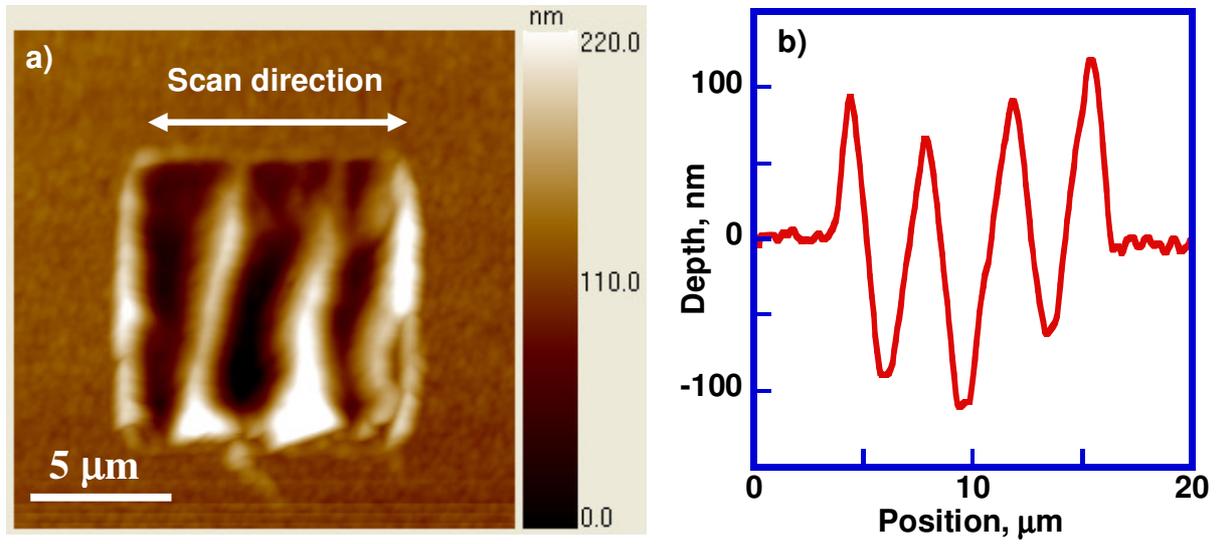

*Fig. 3. a) Wear pattern and b) wear track of a gold film after 200 wear cycles at 10 µN normal load with a sharp Berkovich tip in water showing surface ripples.*

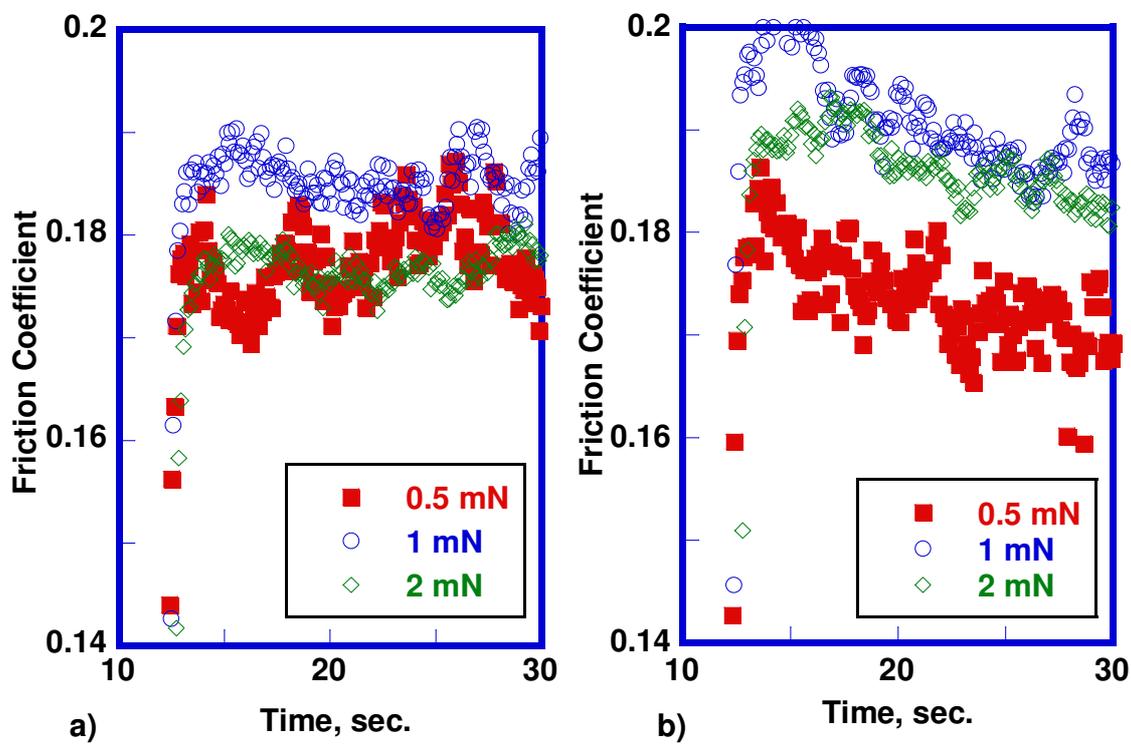

*Fig. 4. Friction coefficient at varying normal load in a) air and in b) water.*

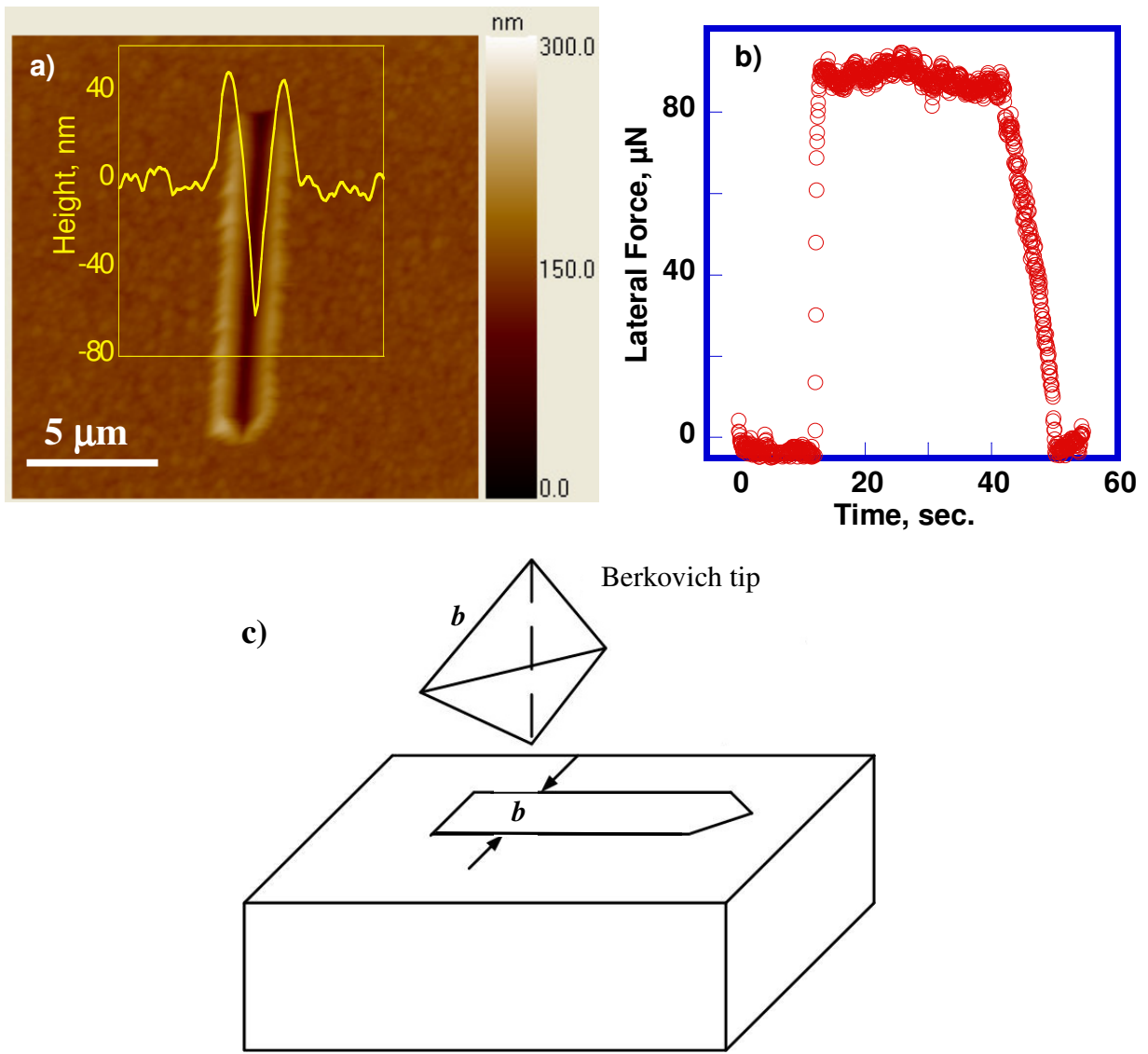

*Fig. 5. a) Scratch topography and b) lateral friction force at 500 µN normal load in air; c) schematic diagram of the scratch test.*

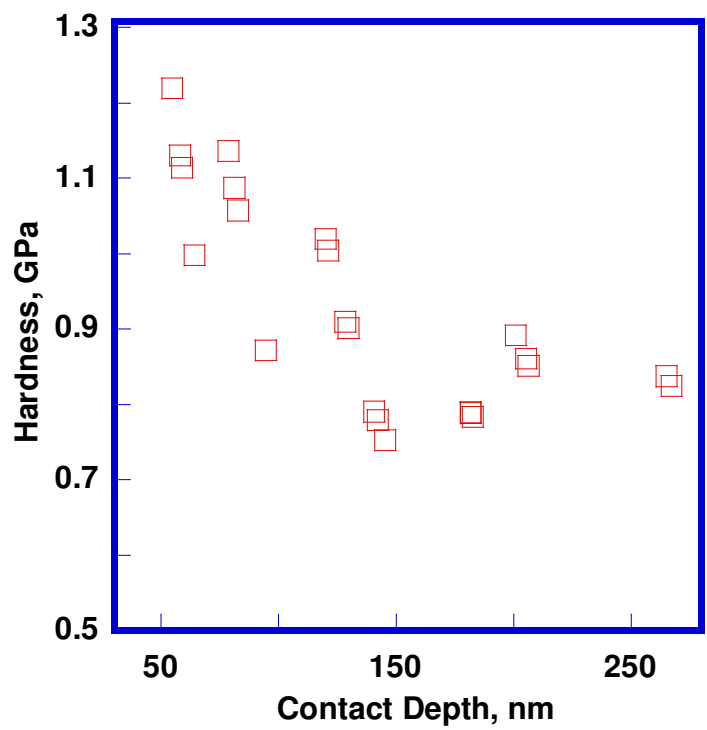

*Fig. 6. Nanoindentation hardness of gold films at different contact depth.*